\documentclass[aps,prc,showpacs,twocolumn,superscriptaddress]{revtex4}
\usepackage{multirow}
\usepackage{graphicx}
\usepackage{amsmath}

\def\version{version 6.4}
\newcommand{\sqrtsNN}{\mbox{$\sqrt{\mathrm{\it s_{NN}}}$} }
\newcommand{\vtwo}{$v_{2}$ }
\newcommand{\vtwos}{$v_{2}$}
\newcommand{\ks}{${K}^{0}_{S}$ }
\newcommand{\kss}{${K}^{0}_{S}$}
\newcommand{\lam}{$\Lambda$ }
\newcommand{\lams}{$\Lambda$}
\newcommand{\xii}{$\Xi$ }
\newcommand{\xiis}{$\Xi$}
\newcommand{\chg}{$h^{\pm}$ }
\newcommand{\chgs}{$h^{\pm}$}
\newcommand{\pt}{$p_T$ }
\newcommand{\pts}{$p_T$}
\newcommand{\ket}{$m_{T} - m$ }
\newcommand{\kets}{$m_{T} - m$}

\newcommand{\nq}{$n_q$ }

\def \auau  {Au+Au }
\def \cucu  {Cu+Cu }
\def \pp    {$p$+$p$ }

\def \npart {$N_{\mathrm{part}}$}

\def \eparttwo {$\varepsilon_{\mathrm{part}}\{2\}$ }
\def \eparttwos {$\varepsilon_{\mathrm{part}}\{2\}$}
\def \GeVc {\mbox{$\mathrm{GeV}/c$}}

\def \ecc {$\varepsilon$ }

\def \llam      {$\Lambda + \overline{\Lambda}$ }

\def \etal   {\mbox{$\mathrm{\it et\ al.}$}}
\def \sims   {$\sim$~}

\usepackage{color}

\begin{document}
\title{
\begin{flushright}
{
\small \sl \version \\
%%%\today \\ for Phys. Rev. C. \\
%\Red{additions} \\
%\Green{subtractions} \\
%\Blue{modifications} \\
%\Magenta{questions} \\
}
\end{flushright}
Charged and strange hadron elliptic flow \\
 in \cucu collisions at \sqrtsNN = 62.4 and 200 GeV }

\affiliation{Argonne National Laboratory, Argonne, Illinois 60439, USA}
\affiliation{University of Birmingham, Birmingham, United Kingdom}
\affiliation{Brookhaven National Laboratory, Upton, New York 11973, USA}
\affiliation{University of California, Berkeley, California 94720, USA}
\affiliation{University of California, Davis, California 95616, USA}
\affiliation{University of California, Los Angeles, California 90095, USA}
\affiliation{Universidade Estadual de Campinas, Sao Paulo, Brazil}
\affiliation{University of Illinois at Chicago, Chicago, Illinois 60607, USA}
\affiliation{Creighton University, Omaha, Nebraska 68178, USA}
\affiliation{Czech Technical University in Prague, FNSPE, Prague, 115 19, Czech Republic}
\affiliation{Nuclear Physics Institute AS CR, 250 68 \v{R}e\v{z}/Prague, Czech Republic}
\affiliation{University of Frankfurt, Frankfurt, Germany}
\affiliation{Institute of Physics, Bhubaneswar 751005, India}
\affiliation{Indian Institute of Technology, Mumbai, India}
\affiliation{Indiana University, Bloomington, Indiana 47408, USA}
\affiliation{Alikhanov Institute for Theoretical and Experimental Physics, Moscow, Russia}
\affiliation{University of Jammu, Jammu 180001, India}
\affiliation{Joint Institute for Nuclear Research, Dubna, 141 980, Russia}
\affiliation{Kent State University, Kent, Ohio 44242, USA}
\affiliation{University of Kentucky, Lexington, Kentucky, 40506-0055, USA}
\affiliation{Institute of Modern Physics, Lanzhou, China}
\affiliation{Lawrence Berkeley National Laboratory, Berkeley, California 94720, USA}
\affiliation{Massachusetts Institute of Technology, Cambridge, MA 02139-4307, USA}
\affiliation{Max-Planck-Institut f\"ur Physik, Munich, Germany}
\affiliation{Michigan State University, East Lansing, Michigan 48824, USA}
\affiliation{Moscow Engineering Physics Institute, Moscow Russia}
\affiliation{City College of New York, New York City, New York 10031, USA}
\affiliation{NIKHEF and Utrecht University, Amsterdam, The Netherlands}
\affiliation{Ohio State University, Columbus, Ohio 43210, USA}
\affiliation{Old Dominion University, Norfolk, VA, 23529, USA}
\affiliation{Panjab University, Chandigarh 160014, India}
\affiliation{Pennsylvania State University, University Park, Pennsylvania 16802, USA}
\affiliation{Institute of High Energy Physics, Protvino, Russia}
\affiliation{Purdue University, West Lafayette, Indiana 47907, USA}
\affiliation{Pusan National University, Pusan, Republic of Korea}
\affiliation{University of Rajasthan, Jaipur 302004, India}
\affiliation{Rice University, Houston, Texas 77251, USA}
\affiliation{Universidade de Sao Paulo, Sao Paulo, Brazil}
\affiliation{University of Science \& Technology of China, Hefei 230026, China}
\affiliation{Shandong University, Jinan, Shandong 250100, China}
\affiliation{Shanghai Institute of Applied Physics, Shanghai 201800, China}
\affiliation{SUBATECH, Nantes, France}
\affiliation{Texas A\&M University, College Station, Texas 77843, USA}
\affiliation{University of Texas, Austin, Texas 78712, USA}
\affiliation{Tsinghua University, Beijing 100084, China}
\affiliation{United States Naval Academy, Annapolis, MD 21402, USA}
\affiliation{Valparaiso University, Valparaiso, Indiana 46383, USA}
\affiliation{Variable Energy Cyclotron Centre, Kolkata 700064, India}
\affiliation{Warsaw University of Technology, Warsaw, Poland}
\affiliation{University of Washington, Seattle, Washington 98195, USA}
\affiliation{Wayne State University, Detroit, Michigan 48201, USA}
\affiliation{Institute of Particle Physics, CCNU (HZNU), Wuhan 430079, China}
\affiliation{Yale University, New Haven, Connecticut 06520, USA}
\affiliation{University of Zagreb, Zagreb, HR-10002, Croatia}

\author{B.~I.~Abelev}\affiliation{University of Illinois at Chicago, Chicago, Illinois 60607, USA}
\author{M.~M.~Aggarwal}\affiliation{Panjab University, Chandigarh 160014, India}
\author{Z.~Ahammed}\affiliation{Variable Energy Cyclotron Centre, Kolkata 700064, India}
\author{A.~V.~Alakhverdyants}\affiliation{Joint Institute for Nuclear Research, Dubna, 141 980, Russia}
\author{I.~Alekseev~~}\affiliation{Alikhanov Institute for Theoretical and Experimental Physics, Moscow, Russia}
\author{B.~D.~Anderson}\affiliation{Kent State University, Kent, Ohio 44242, USA}
\author{D.~Arkhipkin}\affiliation{Brookhaven National Laboratory, Upton, New York 11973, USA}
\author{G.~S.~Averichev}\affiliation{Joint Institute for Nuclear Research, Dubna, 141 980, Russia}
\author{J.~Balewski}\affiliation{Massachusetts Institute of Technology, Cambridge, MA 02139-4307, USA}
\author{L.~S.~Barnby}\affiliation{University of Birmingham, Birmingham, United Kingdom}
% ...

\collaboration{STAR Collaboration}\noaffiliation

\date{\today}

\begin{abstract}
 We present the results of an elliptic flow, \vtwos, analysis of \cucu
collisions recorded with the STAR detector at RHIC at \sqrtsNN $=\ 62.4$ and $200$ GeV.
Elliptic flow as a
function of transverse momentum, \vtwos(\pts), is reported for
different collision centralities for charged hadrons \chgs, and
strangeness containing hadrons \kss, \lams, \xiis, $\phi$ in the midrapidity region
$|\eta| < 1.0$. Significant reduction in systematic uncertainty of
the measurement due to non-flow effects has been achieved by correlating
particles at midrapidity, $|\eta| < 1.0$, with those at forward
rapidity, $2.5 < |\eta| < 4.0$.
We also present azimuthal correlations in \pp collisions at $\sqrt{s}=\ 200$ GeV
to help estimating non-flow effects. To
study the system-size dependence of elliptic flow, we present a
detailed comparison with previously published results from \auau
collisions at \sqrtsNN $=\ 200$ GeV. We observe that \vtwos(\pts) of
strange hadrons has similar scaling properties as were first observed
in \auau collisions, i.e.: (i) at low transverse momenta, $p_T < 2\ \GeVc$,
\vtwo scales with transverse kinetic energy, \kets, and
(ii) at intermediate \pts, $2 < p_T < 4\ \GeVc$, it scales with the
number of constituent quarks, $n_q$. We have found that ideal
hydrodynamic calculations fail to reproduce the centrality
dependence of \vtwos(\pts) for \ks and \lams. Eccentricity
scaled $v_2$ values, $v_{2}/\varepsilon$, are larger in more central collisions, suggesting
stronger collective flow develops in more central
collisions. The comparison with Au+Au collisions which go further in density shows
$v_{2}/\varepsilon$ depend on the system size, number of participants \npart.
This indicates that the ideal hydrodynamic limit is not reached in \cucu
collisions, presumably because the assumption of thermalization is not attained.
\end{abstract}

\maketitle
%--==========================================================================
\clearpage
\section{Introduction}
\label{sect_intro}
At the early stages of relativistic heavy ion collisions, a hot and
dense, strongly interacting medium is created. The subsequent system
evolution is determined by the nature of the medium. Experimentally,
the dynamics of the system evolution has been studied
by measuring the azimuthal anisotropy of the particle production
relative to the reaction
plane~\cite{yingchao98,v2Methods,reviewVoloshin08}.
The centrality of the collision, defined by the transverse distance between the centers of the
colliding nuclei called the impact parameter, results in an
``almond-shaped" overlap region that is spatially azimuthal anisotropic.
It is generally assumed that the initial spatial anisotropy in the system
is converted into momentum-space anisotropy through
re-scatterings~\cite{rqmdv2, sorge97quench}. The elliptic flow, \vtwos, is the second
harmonic coefficient of a Fourier expansion of the final
momentum-space azimuthal anisotropy. Due to the
self-quenching effect, it provides information about
the dynamics at the early
stage of the collisions~\cite{sorge97,ollitrault92,shuryak01}. Elliptic flow can provide information about the
pressure gradients, the effective degrees of freedom, the degree of
thermalization, and equation of state of the matter created at the
early stage. Thus, the centrality and system-size dependence of elliptic
flow at different beam energies can be used to study the properties
of the matter created in heavy ion collisions~\cite{sorge97}.

Recently, two important insights have been obtained from the
experimental results on \vtwos~as a function of transverse
momentum, \pts, in \auau collisions at RHIC.
First, in the low \pt region, $p_T < 2$ GeV$/c$, the hadron mass
hierarchy predicted by ideal hydrodynamic calculations is observed
for identified hadrons $\pi$, $K$, \kss, $p$, \lam and
\xiis~\cite{starwp,phenixwp,starklv2,msv2,staridv2}. Even the $\phi$
and $\Omega$, which are believed to have a reduced cross section for hadronic
interactions~\cite{MSH_section0,MSH_section1,MSH_section2,MSH_section3,MSH_section4,MSH_section5},
are consistent with the mass
ordering~\cite{msv2,star_fv2,phenix_fv2}.
Second, in the intermediate \pt region, $2
< p_T < 4$ GeV$/c$, \vtwos(\pts) follows a scaling depending on the
number of constituent quarks within a given hadron, which can be
explained via coalescence models~\cite{msv2,star_fv2,Coal_Molnar}.
Quark number scaling suggests that the system is in a partonic state
and that the constituent quark degrees of freedom were relevant
during the time \vtwo was developed.

STAR's first published paper showed that elliptic flow at
RHIC is unexpectedly large~\cite{star_130v2}, comparable to
predictions of ideal hydrodynamic calculations~\cite{ollitrault92,idel_hydro1,idel_hydro3,idel_hydro4}.
This observation is among the evidence favoring the picture of a
nearly-perfect liquid~\cite{perfectLiquid}. With the assumption of
thermalization, ideal hydrodynamic calculations predict that the $v_2$
divided by spatial eccentricity, $\varepsilon$, does not depend on
the collision centrality~\cite{centdependence}. The spatial eccentricity is defined by~\cite{ms03}:
\begin{equation} \label{ecc} \varepsilon \ =\
\frac{\langle y^{2} - x^{2}\rangle}{\langle y^{2} + x^{2}\rangle}
\end{equation}
where $x$ and $y$ are the spatial coordinates in the plane
perpendicular to the collision axis. The angle brackets
$\langle\ \rangle$ denote an average weighted by the initial
density.
However, recent RHIC \vtwos/\ecc data for charged hadrons
\chg and strangeness containing hadrons \kss, $\phi$, \lams, \xii show a
trend to increase as a function of the particle density scaled by
the system-size~\cite{STARcum,star_v2cen},
lacking the saturation indicated by ideal hydrodynamic calculations~\cite{star_v2cen}.
This monotonic increase is a feature of a class of model descriptions
that conform to the low-density limit~\cite{ln}.
Whether the thermalization and ideal hydrodynamic limit
are reached or not at RHIC is not conclusive.
A transport model suggested in Ref.~\cite{Drescher07} is constructed to link the low-density limit
to the ideal hydrodynamic limit.
In the microscopic transport picture, the ideal hydrodynamic limit is reached
when the mean free path is very small or the cross section is very large.
With this transport model approach, the degree of thermalization and the ideal hydrodynamic
limit can be addressed~\cite{STARTransport}.

The previous results mainly focus on the centrality dependence of
charged hadron and identified hadron \vtwo in \auau collisions.
Since the conditions in \auau collisions might not hold in smaller
systems and at lower beam energies, the system-size and beam-energy
dependence of identified hadron elliptic flow will shed light on the
properties of partonic collectivity and quark degrees of
freedom. Further, the study of \vtwo in collisions of nuclei smaller than
\auau will allow us to test the early thermalization hypothesis in
\auau collisions. To date, there are only a few studies of identified hadron
\vtwo in \cucu collisions. In this article, we
present the results on the azimuthal anisotropy parameter \vtwos(\pts)  of
\chgs, \kss, \lam, \xii and $\phi$ from \sqrtsNN = 62.4 and 200 GeV
\cucu collisions. As a function of collision centrality, the
scaling properties of \vtwo with the transverse kinetic energy \ket
and the number of constituent quarks \nq are reported.
In the quantity $m_T$ = $\sqrt{p_T^2+m^2}$, $m$
denotes the rest mass of a given hadron.
We also discuss system-size dependence in
this article.

The rest of the paper is organized in the following way:
Section~\ref{sect_analysis} summarizes the analysis details
including data and centrality selections, particle identification
and flow methods used for charged hadrons and identified hadrons. In
the following, we use \chgs, \lam and \xii to denote charged hadron,
\llam and $\Xi^- + \overline{\Xi}^+$, respectively. In
Section~\ref{sect_results}, we present measurements of \vtwo for \chg in
\cucu collisions from different analysis methods. Differences in $v_{2}$
are used to estimate the systematic error.
Section~\ref{sect_discuss} presents the results and physics discussion of the
scaling properties and system-size dependence along with ideal
hydrodynamic calculations. Last, a summary is presented in
Section~\ref{sect_summary}.

\section{Experiments and Analysis}
\label{sect_analysis}

\subsection{Experiments and data sets}

For this article, our data were collected from \sqrtsNN = 62.4 and
200 GeV \cucu collisions with the STAR detector during the  fifth RHIC
run in 2005. In addition
data from \sqrtsNN = 200 GeV \pp collisions in 2005 were used in the
analysis of non-flow contributions. STAR's main
Time Projection Chamber (TPC)~\cite{startpc} and two Forward Time
Projection Chambers (FTPCs)~\cite{starftpc} were used for particle
tracking in the central region ($|\eta| < 1.0$) and forward regions
($2.5 < |\eta| < 4.0$), respectively. Both the TPC and FTPCs provide
azimuthal acceptance over $2\pi$.
Only those events which have
the primary vertex position along the longitudinal beam direction ($V_{z}$)
within 30 cm of the nominal collision point are selected for the analysis. This
is done in order to have a more uniform detector performance within
$|\eta|<1.0$. The centrality definition, which is based
on the raw charged particle TPC multiplicity with $|\eta| < 0.5$,
is the same as used previously~\cite{centralitydef}.
After quality cuts, the number of
the 60\% most central events
is about 24 million for 200 GeV \cucu
collisions and 10 million for 62.4 GeV \cucu collisions. The results from more peripheral
collisions are not presented due to trigger inefficiencies
at low multiplicity.

The centrality was defined using the number of charged tracks
with quality cuts similar to those in Ref.~\cite{star_v2cen}. The 60\% most central events
for \vtwo analysis of \chg were divided into six
centrality bins, each spanning an interval of 10\% of the geometric cross
section. For \vtwo analysis of \ks and $\Lambda$, centrality bins of $0-20\%$ and $20-60\%$
were used.
In order to reduce the multiplicity fluctuations in wide centrality bins, we calculated
\vtwo in the 10\% wide bins, then combined them using
particle yield as the weight.

To select good tracks from primary collisions,
charged particle tracks coming from the collision which transversed the
TPC or FTPCs were selected by requiring the distance of closest
approach to the primary vertex be less than 3 cm. Tracks used for \kss,
\lam and $\Xi$ reconstruction were not subject to this cut.  We required that
the TPC and FTPCs had a number of hits used for reconstruction
of the tracks of the particles $> 15$ and $> 5$, respectively. For
the TPC and FTPCs the ratio of the number of fit hits to maximum
possible hits was $> 0.52$.
An additional transverse momentum cut ($0.15 < p_{T} < 2\ \GeVc$)
was applied to the charged tracks
for the event plane determination.

\subsection{Particle identification}

\begin{figure*}[ht]
\centering \hskip 0cm
\includegraphics[width=1.0\textwidth]{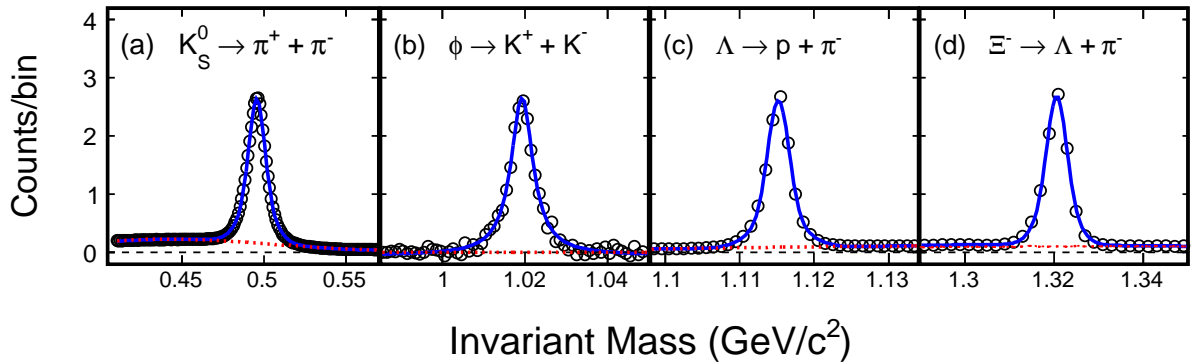}
\caption{ (Color online) Invariant mass distributions for (a) \ks
($1.2 < p_{T} < 1.4\ \GeVc$), (b) $\phi$ ($1.0 < p_{T} < 2.0\ \GeVc$),
(c) $\Lambda$ ($1.4 < p_{T} < 1.6\ \GeVc$) and (d) $\Xi$
($1.25 < p_{T} < 1.75\ \GeVc$) in \sqrtsNN = 200 GeV \cucu
60\% most central collisions. The solid curves represent the fits to the
invariant mass distributions: Gaussians plus fourth-order polynomials for
\kss, $\Lambda$ and $\Xi$, Breit-Wigner plus a linear function for $\phi$. The dotted curves are the estimated
backgrounds: the fourth order polynomials for \ks and $\Lambda$, a linear
function for $\phi$, and a rotation method described in the text for $\Xi$.
For clarity, the invariant mass distributions for \ks, $\Lambda$, $\phi$ and $\Xi$
are scaled by 1/50 000, 1/130 000, 1/5 000 and 1/8 000, respectively.
The error bars are shown only for the statistical uncertainties.} \label{fig_pid}
\end{figure*}

We utilized the topology of decay as measured with the TPC inside the magnetic field to
identify $K_{S}^{0}$, $\phi$, $\Lambda$ ($\overline{\Lambda}$) and $\Xi^{-}$
($\overline{\Xi}^{+}$).
We used the decay channels:
$K_{S}^{0} \rightarrow \pi^{+} + \pi^{-}$,
$\phi \rightarrow$ $K^{+} + K^{-}$,
$\Lambda \rightarrow p + \pi^{-}$
($\overline{\Lambda} \rightarrow \overline{p} + \pi^{+}$),
$\Xi^{-} \rightarrow \Lambda + \pi^{-}$
($\overline{\Xi}^{+} \rightarrow \overline{\Lambda}+ \pi^{+}$).
Similar to the previous analysis in Ref.~\cite{star_v2cen},
topological and kinematic cuts were applied to reduce the
combinatorial backgrounds. Figure~\ref{fig_pid} shows the invariant
mass distributions for (a) $K_{S}^{0}$, (b) $\phi$, (c) $\Lambda$ and
(d) $\Xi$ for selected \pt bins in \sqrtsNN = 200 GeV \cucu
60\% most central collisions. The cuts used for \cucu collisions are
similar to those for \auau collisions in
Ref.~\cite{star_v2cen}.
The combinatorial backgrounds were
estimated from the fourth order polynomial fits for \kss~and~$\Lambda$~\cite{star_v2cen}.
The invariant mass distribution for $\phi$ is
after subtraction of combinatorial background estimated
using event mixing~\cite{mixed_event}, the remaining combinatorial backgrounds were
estimated by a first order polynomial fit~\cite{star_fv2}.
For $\Xi$, the background was estimated by rotating the transverse momentum
of the daughter $\Lambda$ by $180^{0}$. This operation breaks the correlation
between the $\Lambda$ and the other daughter particle. The resulting
invariant mass distributions provide a good approximation of the
true background distribution. The detailed description of the method
can be found in Refs.~\cite{starklv2,msv2}.

\subsection{Flow methods}
\label{sect_method}

Anisotropic flow, which is an anisotropy in the particle production relative to the
reaction plane, results in correlations among particles and can be
studied by the analysis of these correlations. At the same time these
correlations are affected by other effects which are not related to the
orientation of the reaction plane. Such are commonly referred to as
non-flow, and are due, e.g., to resonance decays and  jet production.
Different methods used to measure anisotropic flow are affected by
non-flow effects in different ways, and are used in this analysis
to evaluate the systematic uncertainty of the measurements.

\subsubsection{Event Plane method with TPC event plane}
\label{sect_TPCEP}
The Event Plane method~\cite{v2Methods} uses the anisotropic flow
itself to determine the event plane (the estimated reaction plane),
which can be done for each harmonic. The second harmonic flow vector,
$Q_2$, of the event is constructed using the TPC tracks $i$ in the event
with their
azimuthal angle, $\phi_i$, according to Eq.~(\ref{Q2x}) and
Eq.~(\ref{Q2y}). In order to maximize the resolution of the flow effect, the weights
$w_i$ are set equal to \pt up to 2 \GeVc.
\begin{equation}
\label{Q2x} Q_2\cos(2\Psi_2)\ =\ Q_{2x}\ = \
\sum_{i}w_i\cos(2\phi_i)
\end{equation}

\begin{equation} \label{Q2y} Q_2\sin(2\Psi_2)\ =\ Q_{2y}\ = \
\sum_{i}w_i\sin(2\phi_i)
\end{equation}

Elliptic flow is first calculated with respect to the event plane angle $\Psi_{2}$
as shown in Eq.~(\ref{v2obs}), which is called the observed $v_2$. The angle
brackets indicate an average over all particles in all events. However,
tracks used for the $v_2$ calculation are excluded from the calculation of
the flow vector to remove auto-correlation effects.
Then the observed \vtwo is corrected by the event plane resolution (the
denominator in Eq.~(\ref{v2EP2})) to
obtain $v_2$ relative to the event plane.
\begin{equation} \label{v2obs} v_{2}^{\mathrm{obs}}\ =\ \langle
\cos[2(\phi-\Psi_2)]\rangle
\end{equation}
\begin{equation} \label{v2EP2} v_{2}\ =\
\frac{v_2^{\mathrm{obs}}}{\langle \cos[2(\Psi_2-\Psi_r)]\rangle}
\end{equation}
The results are denoted as $v_{2}\{\rm TPC\}$ in the following.

Since the reaction plane is unknown, the denominator in
Eq.~(\ref{v2EP2}) is still not calculable. As shown in
Eq.~(\ref{EPres}), we estimate the event plane resolution by the
correlations between the azimuthal angles of two subset groups of
tracks, called subevents A and B. In this analysis, we use two
random subevents with equal numbers of particles. In
Eq.~(\ref{EPres}) $C$ is a constant calculated from the known
multiplicity dependence of the resolution~\cite{v2Methods}.
\begin{equation}
\langle \cos[2(\Psi_2-\Psi_r)]\rangle\ =\ C \sqrt { \langle
\cos[2(\Psi_{2}^{A}-\Psi_{2}^{B})] \rangle} \label{EPres}
\end{equation}
In the case
of low resolution ($\leq 0.2$), such as for the FTPC event plane,
$C$ approaches $\sqrt{2}$.

The reaction plane azimuthal distribution should be isotropic in the
laboratory frame. Thus, the event plane azimuthal distribution has to be flat
if the detectors have ideal acceptance. Since the detectors
usually have non-uniform acceptance, a procedure for flattening the event
plane distribution is necessary. For the event plane reconstructed
from TPC tracks, the $\phi$ weight method is an effective way to
flatten the distribution. $\phi$ weights are generated by inverting
the $\phi$ distributions of detected tracks for a large event
sample. The detector acceptance bias is removed by applying the
$\phi$ weight at the $\phi$ of each track to that track. The $\phi$
weights are folded into the weight $w_i$ in Eq.~(\ref{Q2x}) and
Eq.~(\ref{Q2y}). Independent corrections were applied to each
centrality selection in 10\% increments and in 2 bins in the
primary vertex position along the longitudinal beam direction
($V_z$). The corrections were done on a run by run basis
(around $50~k$ events).

\subsubsection{Event Plane method with FTPC event plane}
\label{sect_FTPCEP}
The $\eta$ gap between two FTPCs sitting at two sides of the collision
in the forward regions can be used to reduce non-flow effects due to
short-range correlations. The basic procedures are similar to those
for the Event Plane method with the TPC event plane. There are three
steps: estimate the event plane with FTPC tracks, calculate \vtwo
with respect to the event plane, and obtain the real \vtwo by
correction to the real reaction plane. Equations~(\ref{Q2x})-(\ref{EPres})
can be applied, except that:
i) the sums in Eq.~(\ref{Q2x}) and
Eq.~(\ref{Q2y}) go over FTPC tracks instead of TPC tracks, and
ii) two subset groups of tracks are classified according to the sign of
$\eta$. The tracks with $-4 < \eta < -2.5$ and $2.5 < \eta < 4$ are
called East subevent and West subevent, respectively. Hence, the
resolution in Eq.~(\ref{EPres}) is calculated by the correlation
between the azimuthal angles $\Psi_{2}^{\rm East}$ and $\Psi_{2}^{\rm
West}$. The average in Eq.~(\ref{v2obs}) runs over the TPC tracks as before.
The result of this procedure is denoted as $v_{2}\{\rm FTPC\}$.

Due to the serious loss of acceptance for FTPCs due to
partially non-functioning readout electronics, the number
of tracks detected by the best sector is about 6 times
greater than for the worst one. The result is that the $\phi$ weight method is not
enough to generate a flat event plane distribution.
Thus, further small corrections
are applied after $\phi$ weight corrections using the shift
method~\cite{shiftMethod}. Equation~(\ref{shift}) shows the formula for
the shift correction. The averages in Eq.~(\ref{shift}) are taken from a
large sample of events. In this analysis, the correction is done up
to the twentieth harmonic. This was done in order to make the $\chi^{2}$
divided by the number of degrees of freedom of a flat fit to
the event plane azimuthal angle distribution to be less than 1.
The distributions of $\Psi_{2}^{\rm
East}$ and $\Psi_{2}^{\rm West}$ are separately flattened and then
the full-event event plane distribution is flattened. Accordingly,
the observed \vtwo and resolution are calculated using the shifted
(sub)event plane azimuthal angles.

\begin{equation}
\begin{array}{ll}
\Psi^{'}\  & = \ \Psi \ +\ \sum_{n}\frac{1}{n}[-\langle \sin(2n\Psi)
\rangle \cos(2n\Psi) \\
\\
 & +\ \langle \cos(2n\Psi) \rangle \sin(2n\Psi)]
\end{array}
\label{shift}
\end{equation}

\subsubsection{Scalar Product method}
\label{sect_SPmethod}
The Scalar Product method~\cite{STARcum,runII200gevV2} is similar to
the Event Plane method, and gives \vtwo as:

\begin{equation}
\begin{array}{ll}
v_2(p_T)\ & = \frac{\langle Q_2 u_{2,i}^{*}(p_T) \rangle}{2 \sqrt{
\langle Q_2^{A}Q_2^{B*} \rangle} } \ \\ \\
 \end{array}
\label{scaler}
\end{equation}
where $u_{2,i}=\cos(2\phi_i) + i\sin(2\phi_i)$ is a unit vector of
the $i$th particle, $Q_2=\sum_{k} u_{2,k}$ is the flow vector with the sum
running over all other particles $k$ in the event. The
superscript * denotes the complex conjugate of a complex number. $A$
and $B$ denote the two subevents. In the case that $Q_2$ is
normalized to a unit vector, Eq.~(\ref{scaler}) reduces to the Event
Plane method.
In the Scalar Product method, one can use a
different (re-centering) technique~\cite{recenter}
to correct for detector effects, which presents an
alternative to the weighting and shifting procedures
described in Sec.~\ref{sect_TPCEP}-2 above.
The Scalar Product method is applied to the $v_{2}$ measurement
of charged hadrons.

\subsection{$v_{2}$ versus $m_{inv}$ method}

For $v_2$ of the identified particles $K_{S}^{0}$, $\phi$, $\Lambda$ and $\Xi$,
the $v_{2}$ versus $m_{\mathrm {inv}}$ method is
used~\cite{v2_minv_method, star_v2cen}.
Since $v_2$ is additive, one can write the total
$v_{2}^{\mathrm {Sig+Bg}}$
as a sum of Signal and Background contributions weighted by their
relative yields:
\begin{equation}\begin{aligned}
v_{2}^{\mathrm {Sig+Bg}}(m_{\mathrm {inv}}) = v_{2}^{\mathrm
{Sig}}\frac{\mathrm {Sig}}{\mathrm {Sig+Bg}}(m_{\mathrm {inv}}) +\\
v_{2}^{\mathrm {Bg}}(m_{\mathrm {inv}})\frac{\mathrm {Bg}}{\mathrm
{Sig+Bg}}(m_{\mathrm {inv}}) \label{Equation:v2versusminv}
\end{aligned}\end{equation}
This method involves the calculation of $v_{2}^{\mathrm {Sig+Bg}}$
as a function of $m_{\mathrm {inv}}$ and then fitting the distribution using Eq.~(\ref{Equation:v2versusminv}) with
measured relative yields and parameterizations of $v_{2}^{\mathrm {Sig}}$ and
$v_{2}^{\mathrm {Bg}}$. The $\frac{\mathrm {Bg}}{\mathrm {Sig+Bg}}(m_{\mathrm {inv}})$ distribution is the Bg divided by
(Sig + Bg). The $\frac{\mathrm {Sig}}{\mathrm {Sig+Bg}}(m_{\mathrm {inv}})$ distribution is simply calculated
by $1-\frac{\mathrm {Bg}}{\mathrm {Sig+Bg}}(m_{\mathrm {inv}})$.
The term $v_{2}^{\mathrm {Bg}}$ is parameterized as a
linear function in order to take care of the non-constant $v_{2}^{\mathrm {Bg}}$
value as a function of $m_{\mathrm {inv}}$. The fit result
$v_{2}^{\mathrm {Sig}}$ is the final
observed $v_2$. Why this method works well for measuring signal $v_2$
is explained as follows: a set of data points is used in the fit
over a wide $m_{\mathrm {inv}}$ region for Sig and Bg. Data points far from
the mass peak constrain $v_{2}^{\mathrm {Bg}}(m_{\mathrm {inv}})$,
since pure Bg is expected
in this region. Under the
peak, the $v_{2}^{\mathrm {Sig+Bg}}(m_{\mathrm {inv}})$
is dominated by the Sig distribution.
Finally, the $v_2$ signal is extracted by the fitting method shown
in Eq.~(\ref{Equation:v2versusminv}).

Note that the subtraction procedure used to extract the $v_2$
signal for a given identified particle is independent of the
flow correlations. The $v_2$ distributions of the overall signal and
background are evaluated by one of the flow analysis methods discussed in Sec.~\ref{sect_TPCEP}-3.
In this paper, the Event Plane method with the FTPC event plane is applied for $K_{S}^{0}$,
$\phi$, $\Lambda$ and $\Xi$.

\begin{figure}[t]
\vskip 0cm
\includegraphics[width=0.4\textwidth]{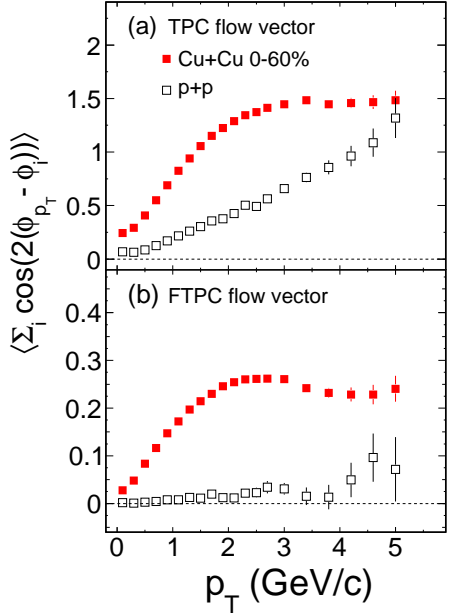}
\caption{(Color online) Charged hadron azimuthal correlations as a
function of \pt in \sqrtsNN = 200 GeV 60\% most central \cucu
collisions (closed squares) compared to those from \sqrtsNN = 200 GeV
\pp collisions (open squares).
Flow vector calculated from (a) TPC
tracks, (b) FTPC tracks. The error bars are shown only for the statistical uncertainties.} \label{Plot::AA_pp}
\end{figure}

\begin{figure}[t]
\vskip 0cm
%\hskip -3.5cm
\centering{\includegraphics[width=0.4\textwidth]{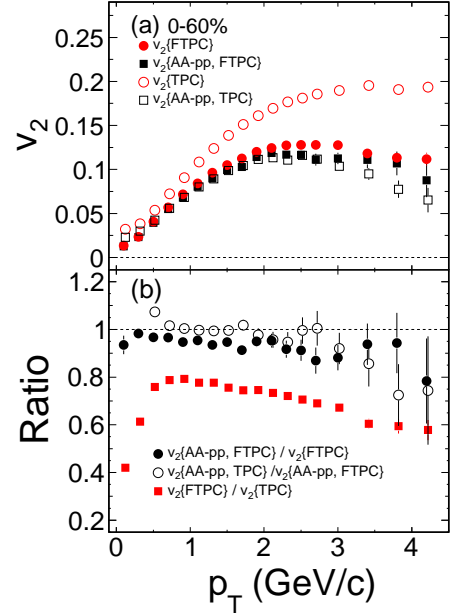}}
\vskip 0.5cm \caption{ (Color online) (a) Charged hadron
\vtwos(\pts) in \sqrtsNN = 200 GeV $0-60\%$ \cucu collisions. Open
circles, closed circles, open squares and closed squares represent the
results of $v_{2}$ as function of \pt measured by the TPC flow
vector ($v_{2}\{\rm TPC\}$), the FTPC flow vector ($v_{2}\{\rm
FTPC\}$), the TPC and FTPC flow vector with subtracting the
azimuthal correlations in \pp collisions ($v_{2}\{AA-pp, \rm TPC\}$,
$v_{2}\{AA-pp, \rm FTPC\}$). (b) The ratio of the results for the
various methods described in (a).
The error bars are shown only for the statistical uncertainties.} \label{v2chgmini}
\end{figure}
\subsection{Non-flow contribution for various methods}

\begin{figure*}[ht]
\hskip -2.0cm
\includegraphics[width=0.6\textwidth]{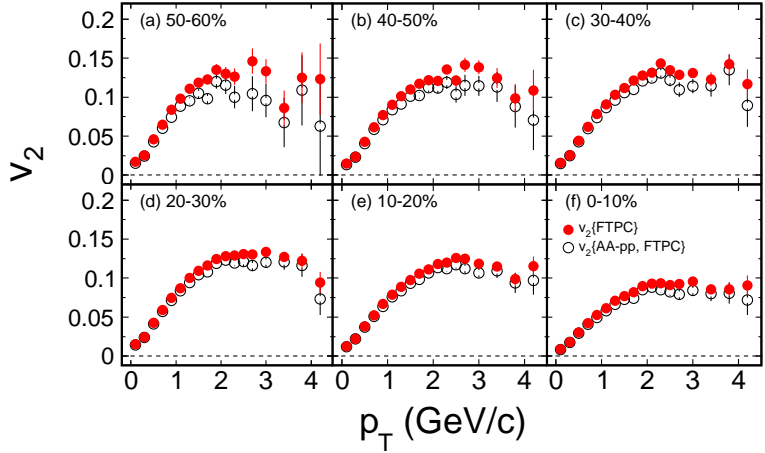}
\caption{ (Color online) Charged hadron $v_{2}\{\rm FTPC\}$ (closed
circles) and $v_{2}\{AA-pp, \rm FTPC\}$ (open circles) as function
of \pt in \sqrtsNN = 200 GeV \cucu collisions for centrality bins:
(a) $50-60\%$, (b) $40-50\%$, (c) $30-40\%$, (d) $20-30\%$, (e) $10-20\%$ and
(f) $0-10\%$. The percentages refer to fraction of
most central events. The error bars are shown only for the statistical uncertainties.} \label{v2chgcen}
\end{figure*}

\begin{figure*}[ht]
\hskip -2.0cm
\includegraphics[width=0.6\textwidth]{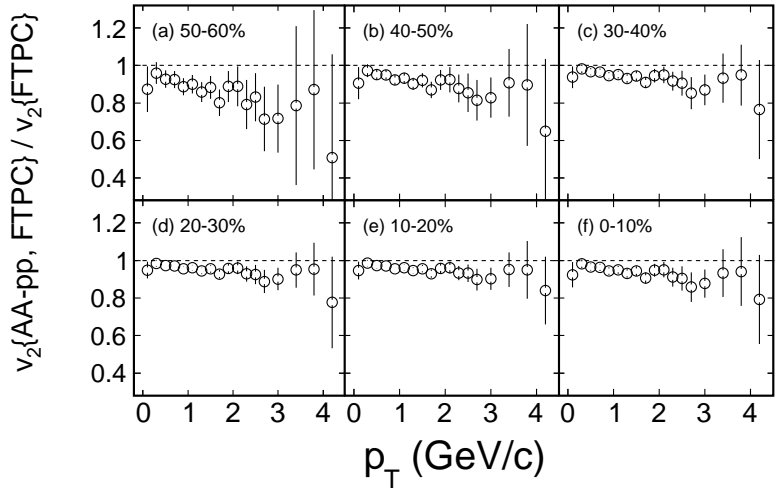}
\caption{Ratios of $v_{2}\{AA-pp,
\rm FTPC\}$/$v_{2}\{\rm FTPC\}$ for charged hadron as function of \pt
in \sqrtsNN = 200 GeV \cucu collisions for centrality bins: (a)
$50-60\%$, (b) $40-50\%$, (c) $30-40\%$, (d) $20-30\%$, (e) $10-20\%$ and (f)
$0-10\%$. The percentages refer to fraction of
most central events.} \label{v2Ratiochgcen}
\end{figure*}

\begin{figure*}[ht]
\includegraphics[width=0.65\textwidth]{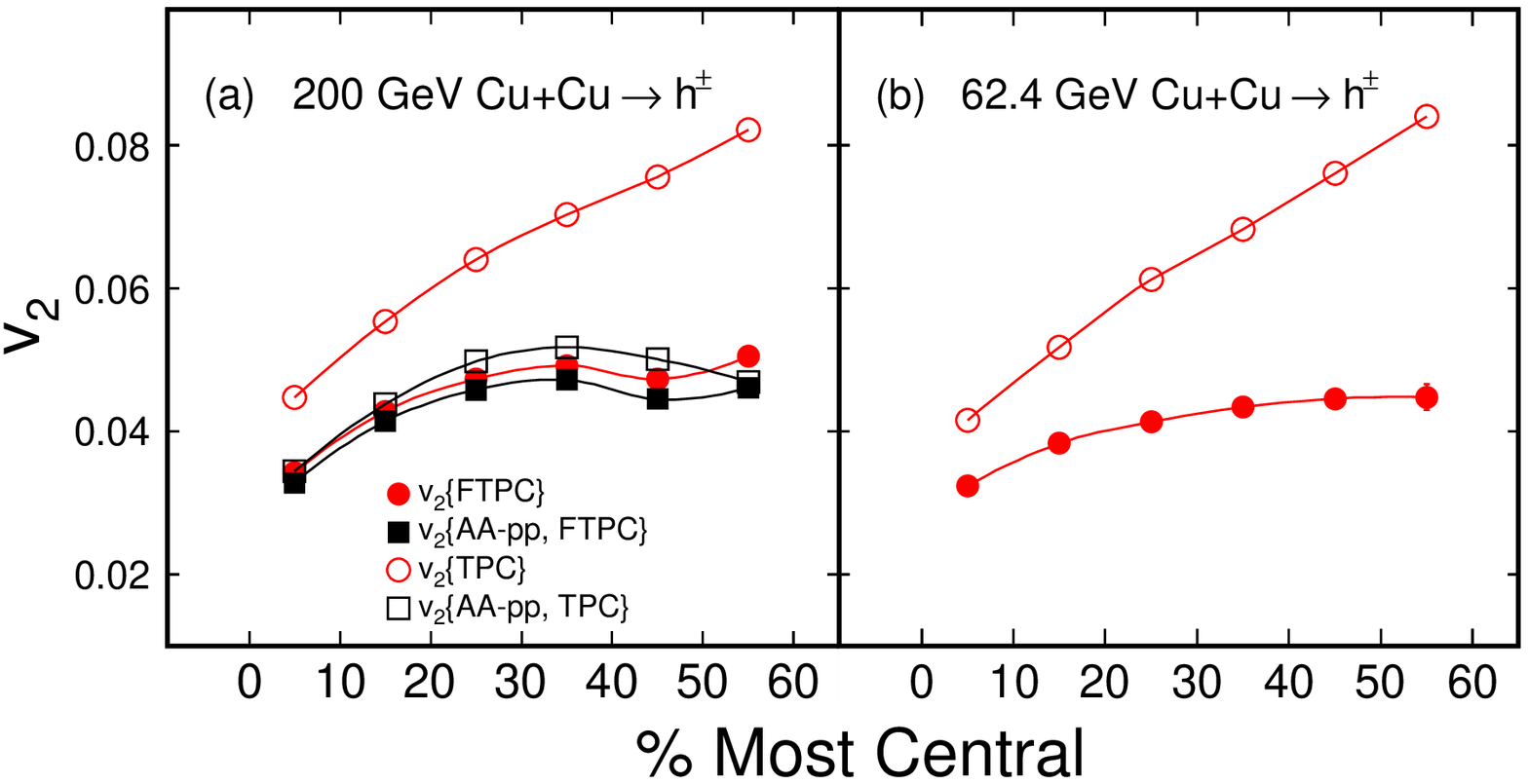}
\caption{ (Color online) Charged hadron \vtwo integrated over \pt
and $\eta$ vs. centrality for the various methods described in the
text in \sqrtsNN = 200 GeV and 62.4 GeV \cucu collisions. The error bars are shown only for the statistical uncertainties.}
\label{Integral}
\end{figure*}

The method of determining \vtwo using cumulants of various orders
has been shown to eliminate non-flow correlations.
However, the method is useful only for large values of flow and multiplicity.
For the relatively low values of flow and multiplicity seen in \cucu collision,
the non-flow correlations have been estimated, as described below.

The Event Plane method with the TPC event plane is sensitive to non-flow
effects. Particles of interest tend to correlate with particles
used in the flow vector calculation due to short-range non-flow
correlations. Also, particles of two random sub-events tend to have
those correlations. Thus, non-flow exists in both the observed \vtwo
(Eq.~(\ref{v2obs})) and the resolution
(Eq.~(\ref{EPres})). To reduce non-flow effects due to short-range
correlations, we take advantage of the large $\eta$ gap between the two
FTPCs sitting at the two sides of the collision in the forward regions.
Non-flow is reduced by the $\eta$ gap
between the TPC and FTPCs, but this may
not be large enough to remove all non-flow correlations. Thus, we
investigate these effects by comparing the azimuthal correlations
measured in \cucu to those in \pp collisions, where all correlations
are assumed to be of non-flow origin~\cite{AA_pp}. Taking into
account the non-flow contribution, the numerator of
Eq.~(\ref{scaler}) can be written as follows~\cite{AA_pp, STARcum}:
\begin{equation}\begin{aligned}
\langle Q_2 u_{2,i}^{*}(p_T) \rangle
&= \left\langle\sum_{k}\cos[2(\phi_{p_{T}} - \phi_{k})]\right\rangle \\
&= Mv_{2}(p_{T})\overline{v_{2}}\ +\ \rm {nonflow}
\label{Equation:scalar}
\end{aligned}
\end{equation}
where $\phi_{p_{T}}$ is the azimuthal angle of particles from a
given $p_{T}$ bin ($u_{2,i}^{*}$ in Eq.~(\ref{scaler})) and
the sum goes over all tracks $k$ in an
event used to determine the flow vector ($Q_{2}$ in Eq.~(\ref{scaler})).
The angled brackets denote averaging over the events.
The first term in the right-hand side of Eq.~(\ref{Equation:scalar})
represents the contribution from elliptic flow. $v_{2}(p_{T})$ is
the value of elliptic flow at a given $p_{T}$. $\overline{v_{2}}$ is
the elliptic flow on average for all particles used in the sum of Eq.~(\ref{Equation:scalar}).
The multiplicity of particles
contributing to the sum is denoted by $M$.
All other
correlations subject to non-flow go to the second term in the
right-hand side of Eq.~(\ref{Equation:scalar}).
It is assumed that the quantity $\langle Q_2 u_{2,i}^{*}(p_T) \rangle$ in \pp collisions
can be used to estimate the non-flow in $AA$ collisions~\cite{AA_pp, runII200gevV2}.
\begin{equation}\begin{aligned}
Mv_{2}(p_{T})\overline{v_{2}}\
=  \langle Q_2 u_{2,i}^{*}(p_T) \rangle_{AA} - \ \langle Q_2 u_{2,i}^{*}(p_T) \rangle_{pp}
\label{Equation:assumption}
\end{aligned}
\end{equation}
%Subtracting the non-flow contributions in the numerator of
%Eq.~(\ref{scaler}) based
%on \pp collisions, $v_{2}\{AA-pp\}$ gives
Dividing both sides by $2 \sqrt{\langle Q_2^{A}Q_2^{B*} \rangle_{AA}}$ as in Eq.~(\ref{scaler}) gives
\begin{equation}
\begin{array}{ll}
v_2\{AA-pp\}(p_T)\ =  & \frac{\langle Q_2 u_{2,i}^{*}(p_T)
\rangle_{AA} - \langle Q_2 u_{2,i}^{*}(p_T) \rangle_{pp}}{2 \sqrt{
\langle Q_2^{A}Q_2^{B*} \rangle_{AA}} }
\end{array}
\label{scalerAApp}
\end{equation}
because  $2 \sqrt{\langle Q_2^{A}Q_2^{B*} \rangle_{AA}}= 2 \sqrt{(M/2) \overline{v_{2}} (M/2) \overline{v_{2}}} = M \overline{v_{2}}$.

Comparing \pp and $AA$ collisions, one might expect
some changes in particle correlations: there could be an
increase in correlations due to a possible increase of jet
multiplicities in $AA$ collisions or, conversely, some
decrease due to the suppression of high $p_{T}$ back-to-back
correlations~\cite{backtoback}.
On the other hand, $AA$ collisions exhibit long $\eta$ range correlations (the ``ridge")~\cite{longetacorr1, longetacorr2},
which are not seen in \pp collisions and the origin of which
is under investigation~\cite{starridge}.
Thus it is difficult to make an accurate estimate of non-flow contributions.
The fact that at high $p_{T}$ ($p_{T}>5\ \GeVc$) the \pp results are very close to central
\auau~\cite{AA_pp, runII200gevV2} suggests that the uncertainties
are relatively small.
In the following we estimate the systematic uncertainties arising from non-flow
contributions. We use $v_{2}\{AA-pp, \rm TPC\}$ and $v_{2}\{AA-pp, \rm FTPC\}$
to denote $v_{2}\{AA-pp\}$ calculated with TPC and FTPC flow vectors, respectively.

\section{Systematic uncertainties}
\label{sect_results}
\begin{table}[t]
\centering \vskip 1.5cm
\caption{Summary of systematic errors of $v_{2}$ due to the reconstruction
procedure of strange hadrons in \cucu~collisions at \sqrtsNN~= 200 GeV.}
\begin{tabular}{ccccc} \hline\hline
     &\multicolumn{2}{c}{$K_{S}^{0}$}&\multicolumn{2}{c}{\lam}\\ \hline
      Centrality&Background&Cut criteria&Background&Cut criteria\\ \hline
     \ $0-60$\%\  &$1\%$& $2\%$&$1\%$& $2\%$\\
     \ $0-20$\%\  &$1\%$& $2\%$&$1\%$& $4\%$\\
     \ $20-60$\%\  &$4\%$& $1\%$&$5\%$& $1\%$\\ \hline
     \hline
\end{tabular}

\label{tab:syserrorsbg}
\end{table}

Non-flow is one of the largest uncertainties in elliptic flow
measurements. As we mentioned in
Section~\ref{sect_method}, this effect can be investigated by
comparing the azimuthal correlations measured in \cucu collisions to
those in \pp collisions. The event average of the sum of the
correlations is given by Eq.~(\ref{Equation:scalar}).

Figure~\ref{Plot::AA_pp} shows the azimuthal correlation,
Eq.~(\ref{Equation:scalar}), as a function of \pt for the $0-60\%$
centrality range in \cucu
collisions at \sqrtsNN~= 200 GeV, compared to \pp
collisions. As we can see, the azimuthal
correlations in \cucu collisions, shown as solid squares, increase
with \pt and then saturate above 2 \GeVc~while those in \pp
collisions, shown as open squares, monotonically increase with $p_T$
in the case of the TPC flow vector. With the flow vector determined from
FTPC tracks the azimuthal
correlations around midrapidity in \pp collisions are small when \pt is less than 4
\GeVc. It means that
one strongly reduces the non-flow effects with the FTPC flow vector
relative to the one seen with the TPC flow vector.

\begin{figure*}[ht]
\includegraphics[width=0.65\textwidth]{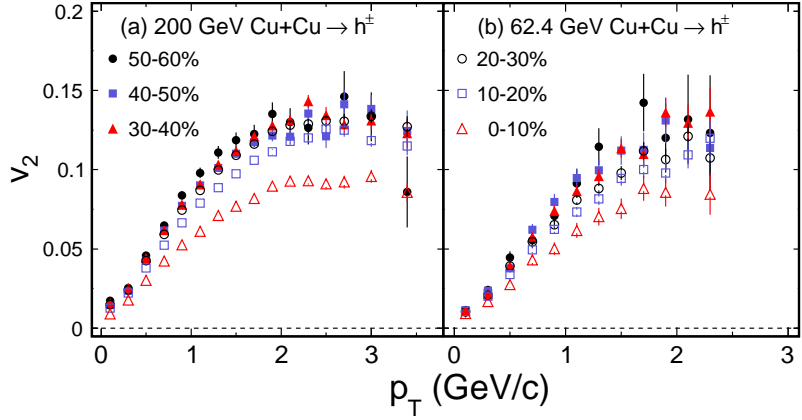}
\caption{ (Color online) Charged hadron $v_{2}$ as function of \pt
for $50-60\%$ (solid circles), $40-50\%$ (solid squares), $30-40\%$ (solid
triangles), $20-30\%$ (open circles), $10-20\%$ (open squares) and
$0-10\%$ (open triangles) in \sqrtsNN = 200 GeV and 62.4 GeV \cucu
collisions. The error bars are shown only for the statistical uncertainties.} \label{v2chgpt}
\end{figure*}

\begin{figure}[t]
\vskip 0cm
\centering\includegraphics[width=0.4\textwidth]{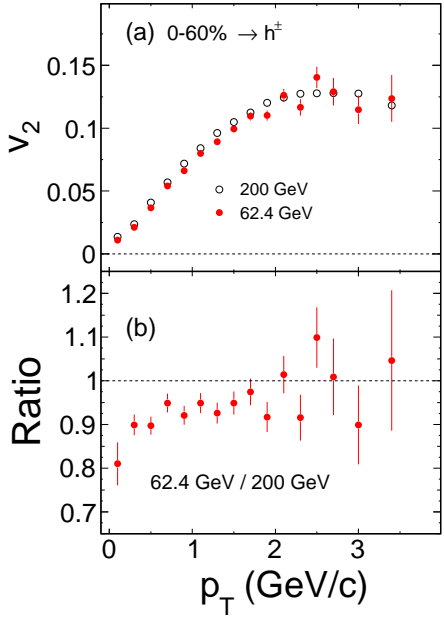}
\caption{ (Color online) (a) Charged hadron $v_2$ as a function of
\pt in \cucu collisions. The results from \sqrtsNN = 200 GeV and
62.4 GeV are presented by open symbols and closed symbols,
respectively. (b) Ratios of the $v_2$($p_T$) from \sqrtsNN = 62.4 GeV
to 200 GeV. The error bars are shown only for the statistical uncertainties.} \label{v2energy}
\end{figure}

\begin{figure*}[t]
\includegraphics[width=1\textwidth]{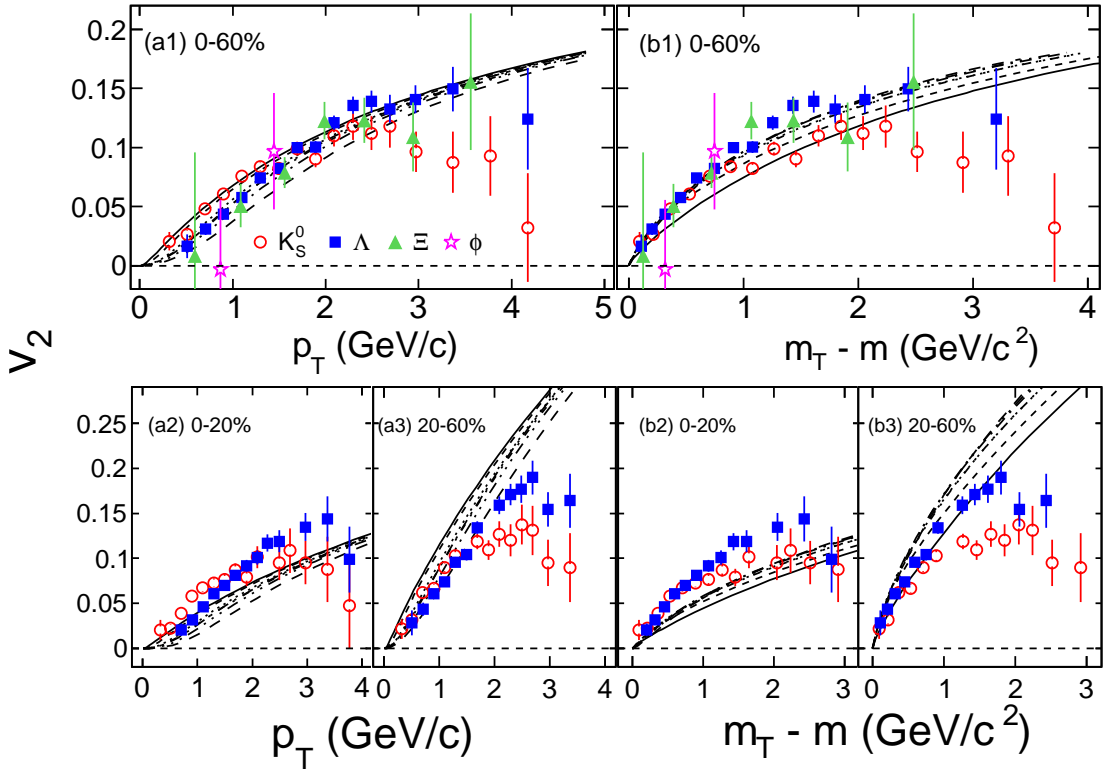}
\caption{(Color online) \vtwo of \ks (open circles), $\Lambda$ (
solid squares), $\Xi$ (solid triangles) and $\phi$ (open stars) as a
function of \pt for (a1) $0-60\%$, (a2) $0-20\%$, (a3) $20-60\%$ and as a
function of \ket~for (b1) $0-60\%$, (b2) $0-20\%$, (b3) $20-60\%$. For
comparisons, the results from ideal hydrodynamic
calculations~\cite{pasirev06,pasi08} are also shown. At a given
$p_T$, from top to bottom, the curves represent $\pi$, $K$, $p$,
$\phi$, $\Lambda$, $\Xi$ and $\Omega$. When \pt is converted to
\kets, this mass hierarchy is reversed in the model results. All
data are from \sqrtsNN = 200 GeV \cucu collisions. The error bars are shown only for the statistical uncertainties.} \label{v2PtMt}
\end{figure*}

\begin{figure*}[t]
\includegraphics[width=0.8\textwidth]{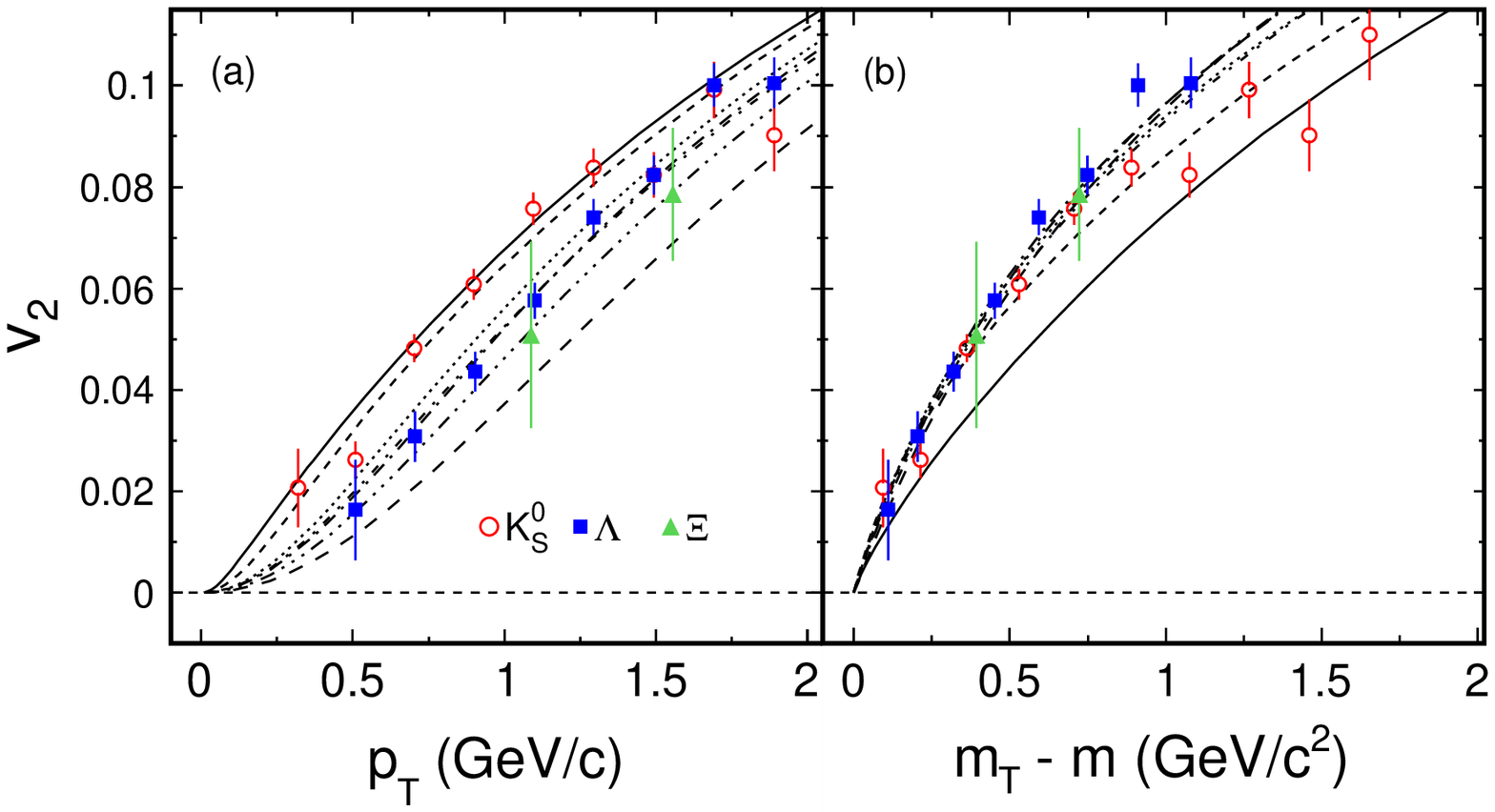}
\caption{(Color online) Same as Fig.~\ref{v2PtMt}~(a1) and (b1), but expanded
for the low \pts~and \ket regions. The data points with large errors have not been plotted.
At a given
$p_T$, from top to bottom, the curves represent the ideal hydrodynamic calculations for
$\pi$, $K$, $p$,
$\phi$, $\Lambda$, $\Xi$ and $\Omega$~\cite{pasirev06,pasi08}. When \pt is converted to
\kets, this mass hierarchy is reversed in the model results.
All data are from $0-60\%$ \cucu collisions at \sqrtsNN = 200 GeV.
The error bars are shown only for the statistical uncertainties.} \label{v2PtMtZoomin}
\end{figure*}

In order to illustrate the sensitivity to non-flow for the various flow analysis methods,
we first analyzed \chg elliptic
flow in the 60\% most central \cucu collisions at \sqrtsNN = 200
GeV. As shown in Fig.~\ref{v2chgmini}~(a),
the fact that $v_{2}\{\rm TPC\}$ is significantly larger than
$v_{2}\{\rm FTPC\}$ indicates a larger non-flow effect in $v_{2}\{\rm
TPC\}$. With the large $\eta$ gap between West and East FTPCs,
non-flow effects due to the short-range correlations are reduced in
$v_{2}\{\rm FTPC\}$. $v_{2}\{\rm FTPC\}$ saturates at \pt \sims 2.5
\GeVc~and then falls off slightly up to \pt \sims 4 \GeVc. In order to
estimate the remaining non-flow effects in $v_{2}\{\rm FTPC\}$,
we subtract the azimuthal correlations of \pp collisions from those in \cucu collisions
according to Eq.~(\ref{scalerAApp}). In
Fig.~\ref{v2chgmini}~(a), $v_{2}\{AA-pp,\rm FTPC\}$ is close to
$v_{2}\{\rm FTPC\}$ in the region $p_{T}<4~\GeVc$.
To quantitatively illustrate non-flow
systematic uncertainties, Fig.~\ref{v2chgmini}~(b) shows the ratios
of $v_{2}\{AA-pp,\rm FTPC\}$ to $v_{2}\{\rm FTPC\}$,
$v_{2}\{AA-pp,\rm TPC\}$ to $v_{2}\{AA-pp,\rm FTPC\}$
and $v_{2}\{\rm FTPC\}$ to $v_{2}\{\rm TPC\}$ as a
function of $p_T$. $v_{2}\{\rm FTPC\}$/$v_{2}\{\rm TPC\}$ shows that
non-flow in $v_{2}\{\rm TPC\}$ increases from 20\% at \pt \sims 0.8
\GeVc~to 40\% at \pt \sims 3.5 \GeVc.
Based on the comparison between $v_{2}\{AA-pp,\rm FTPC\}$ and $v_{2}\{\rm FTPC\}$,
the residual non-flow in $v_{2}\{\rm FTPC\}$ is less than 10\% below \pt \sims 4
\GeVc. We also checked the $v_{2}\{AA-pp\}$ calculated with the TPC flow vector.
Beyond $p_{T}\sim3\ \GeVc$, $v_{2}\{AA-pp,\rm TPC\}$ seems systematically lower,
but within errors it is similar to $v_{2}\{AA-pp,\rm FTPC\}$.
This shows that most of the non-flow is eliminated by subtracting the
azimuthal correlation in \pp collisions, validating our
earlier assumption.

To illustrate the centrality dependence of the systematic
uncertainties, Fig.~\ref{v2chgcen} shows $v_{2}\{\rm FTPC\}$
and $v_{2}\{AA-pp,\rm FTPC\}$ as a function of \pt
for six centrality bins. Ratios of $v_{2}\{AA-pp,\rm FTPC\}$ to $v_{2}\{\rm
FTPC\}$ for each centrality bin are shown in
Fig.~\ref{v2Ratiochgcen} from (a) the most peripheral bin $50-60\%$ to
(f) the most central bin $0-10\%$. For each centrality bin, the ratio
falls off slightly as \pt increases. For the two peripheral bins $50-60\%$ and
$40-50\%$, the ratios drop faster than in the other bins, indicating larger
non-flow contributions in $v_{2}\{\rm FTPC\}$($p_T$) in peripheral
\cucu collisions. Figure~\ref{Integral} shows charged hadron \vtwo
integrated over \pt ($0.15< p_{T}<4\ \GeVc$) and $\eta$
($|\eta| < 1.0$) vs. centrality for the various methods.
It is clear that $v_{2}\{\rm TPC\}$ is much higher than for the other methods,
especially for the peripheral collisions.

To summarize the non-flow systematics we employed the Scalar Product
method with TPC and FTPC flow vectors
for \chg in \cucu collisions at \sqrtsNN = 200 GeV. The results for
the 60\% most central events are shown in Fig.~\ref{v2chgmini}. $v_{2}\{\rm
TPC\}$ has large non-flow contributions while $v_{2}\{\rm FTPC\}$ eliminates most of
the non-flow. In what follows, we will report our results in term of
$v_{2}\{\rm FTPC\}$. For simplicity $v_{2}$ denotes $v_{2}\{\rm
FTPC\}$ except when the flow method is explicitly specified. With
the assumption of pure non-flow effects in \pp collisions, we use
$v_{2}\{AA-pp,\rm FTPC\}$ to estimate non-flow systematic errors in
$v_{2}\{\rm FTPC\}$. Ratios of $v_{2}\{AA-pp,\rm FTPC\}$ to $v_{2}\{\rm FTPC\}$
are shown for the 60\% most central events in Fig.~\ref{v2chgmini}~(b) and six
centrality bins in Fig.~\ref{v2Ratiochgcen}. The ratios show that
non-flow effects increase with \pt for all centrality bins and non-flow effects are larger
in more peripheral bins. In order to estimate the non-flow systematic
error in $v_{2}\{\rm FTPC\}$, we fitted a constant to the ratio
$v_{2}\{AA-pp,\rm FTPC\}$/$v_{2}\{\rm FTPC\}$ in the $p_{T}$ range (0, 4 GeV).
We take the numerical value of this constant as the estimate
     of the systematic uncertainty.
The resulting non-flow systematic error is minus $5\%$ for $0-10\%$, $10-20\%$,
$20-30\%$ and $30-40\%$
collisions and minus $10\%$ for $40-50\%$ and $50-60\%$ collisions.
Although for $K_{S}^{0}$, $\phi$, $\Lambda$ and $\Xi$
$v_{2}$ the non-flow effects may be different, as we don't have enough statistics to repeat the analysis,
we simply assume a similar magnitude for the non-flow systematic error.

The other systematic uncertainties in the $v_2$ analysis procedure are
studied as follows. We estimate the systematic errors from the shifting method
for the FTPC event plane by comparing \vtwo using different maximum harmonics in
Eq.~(\ref{shift}) and find the systematic errors
are less than 1\%. The systematic errors in \ks and \lam \vtwo
resulting from the background uncertainty and topological cut
criteria are estimated using the Event Plane method. The uncertainty
due to the background subtraction is estimated as the relative differences in $v_{2}$ from
fitting the background using second and fourth order polynomials.
The systematic uncertainty is also estimated by varying the cut parameters.
The systematic errors for \ks and $\Lambda$ from the
background uncertainty and the cut criteria are summarized in
Table~\ref{tab:syserrorsbg}. From Ref.~\cite{star_v2cen}, the estimated
systematic uncertainty of $\Lambda$ from feed-down is
less than $2\%$.

\section{Results and Discussion}
\label{sect_discuss}
\subsection{Charged hadrons}

Flow results for charged hadrons
were determined using the Scalar Product method Eq.(8)
with the flow vector derived from the FTPC tracks. A
comparison to $v_{2}\{AA-pp,\rm FTPC\}$ Eq.~(\ref{scalerAApp}) was used to
estimate the systematic error.
Figure~\ref{v2chgpt} shows $v_2$($p_T$) of \chg for six centrality
bins from \cucu collisions at \sqrtsNN = 200 and 62.4 GeV. For a
given centrality bin, $v_2$($p_T$) initially increases with $p_T$.
At higher \pts~($p_T > 2$ GeV$/c$), \vtwo appears to saturate or
decrease. \vtwos(\pts) in more peripheral collisions increases
faster and reaches higher values as expected for the larger eccentricity.

 At low $p_T$, the increase of $v_2$($p_T$) with \pt is
consistent with predictions from ideal hydrodynamic calculations,
which will be shown in Fig.~\ref{v2PtMt} for identified particles.
The model predicts that \vtwo continues increasing beyond \pt $\sim$ 2 \GeVc.
The observed saturation or decrease of $v_2$($p_T$) indicates
that the model is not valid in this region. One expects that
the model should be valid up to higher \pt in a system with
larger densities and larger volumes. This was observed in 200 GeV \auau
collisions~\cite{yutingsqm08} where $v_2$($p_T$) of \chg saturated at
higher \pt in more central collisions. However, we do not observe
the strong centrality dependence of saturation \pt for 200 GeV
\cucu collisions.

Figure~\ref{v2energy} shows the comparison of \vtwo for \chg from
\sqrtsNN = 62.4 and 200 GeV \cucu collisions. The $p_{T}$ dependence of
\vtwo at the two energies is similar.

\subsection{Identified hadrons}
\begin{figure*}[t]
\includegraphics[width=1\textwidth]{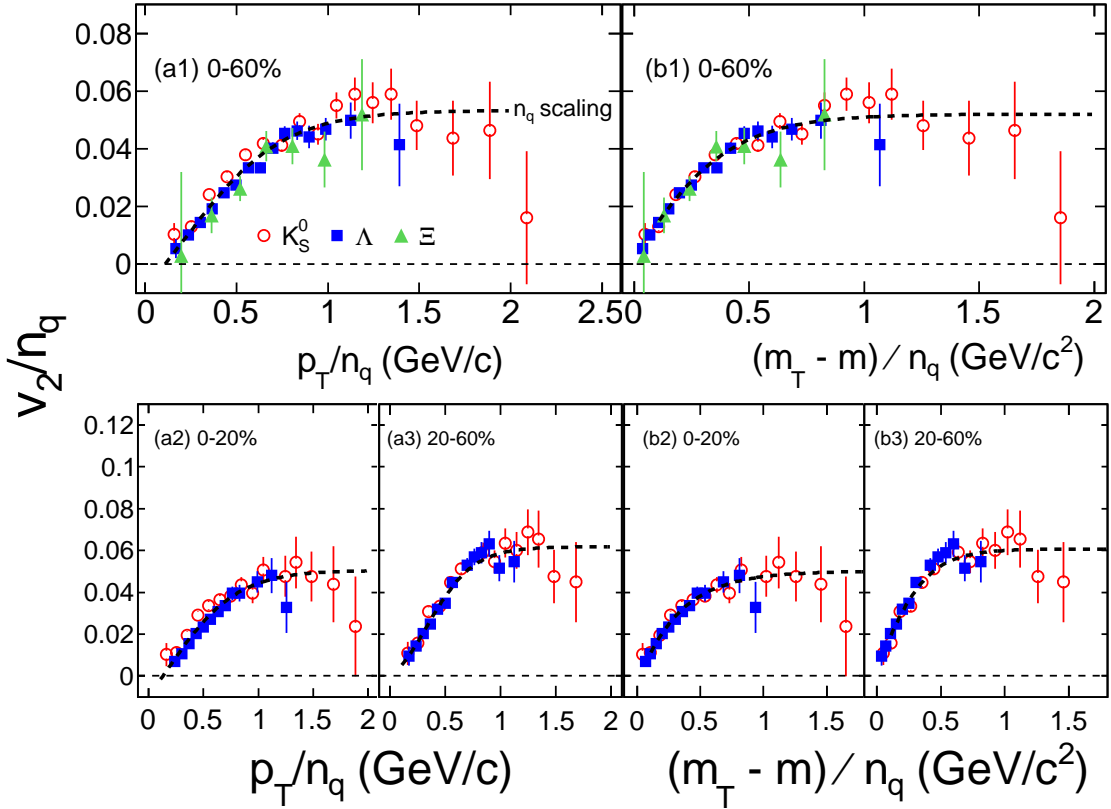}
\caption{(Color online)
\vtwo/\nq versus \pts/\nq (panels (a1)-(a3)) and (\kets)/\nq (panels
(b1)-(b3)), where \nq is the number of constituent quarks in the hadron.
 The parametrization Eq.~(\ref{Equation:Fitv2}) fitted
to the data is shown as the dashed curves. All
data are from \sqrtsNN = 200 GeV \cucu collisions. The error bars are shown only for the statistical uncertainties.} \label{v2NqPtMt}
\end{figure*}

The Event Plane method Eqs.~(\ref{Q2x})-(\ref{v2EP2}) with event plane determined
from the FTPC tracks was applied to ${K}^{0}_{S}$, $\Lambda$, $\Xi$ and $\phi$.
The results are shown in Fig.~\ref{v2PtMt} for the 60\% most central events
and also for the $0-20\%$ and $20-60\%$ centrality bins at
midrapidity $|y|$ $<$ 1.
Due to limited statistics, $\Xi$ and $\phi$
are only shown for the $0-60\%$ bin. The results from
ideal hydrodynamic calculations~\cite{pasirev06,pasi08} for each
centrality bin are shown for $\pi$, $K$, $p$, $\phi$, $\Lambda$,
$\Xi$ and $\Omega$, which are displayed by the curves
from top (bottom) to bottom (top) for the $p_{T}$ (\kets)
dependence.

The gross features of \pt dependence and hadron-type dependence are
similar to those observed in 200 GeV \auau
collisions~\cite{star_v2cen}. At low \pts, the hadron mass
hierarchy (at a given $p_T$, the heavier the hadron, the smaller
$v_2$($p_T$)) is reproduced by ideal hydrodynamic calculations.
(See Fig.~\ref{v2PtMtZoomin}~(a) for clarity.)
Multi strange-quark hadrons $\Xi$ and $\phi$, which
participate less in later hadronic interactions than do
single strange-quark hadrons \ks and $\Lambda$, have sizable \vtwos. In
particular, $\Xi$ is consistent with the mass ordering shown in
Fig.~\ref{v2PtMt}~(a1).
While the model can roughly reproduce the magnitude of the data
for the 60\% most central events sample, there
is an obvious disagreement in centrality selected data.
The model under-predicts
$v_2$($p_T$) in the $0-20\%$ bin while it over-predicts the data in
the $20-60\%$ bin. Effects not included in the model which may be
relevant are geometrical fluctuations in the initial conditions and finite
viscosity effects. It is unclear whether these effects can account
for the difference between the model and data.

At higher $p_T$, the hydrodynamic model mass ordering breaks. $v_2$($p_T$)
appears to depend on hadron type: $v_2$($p_T$) is grouped by mesons
and baryons, with magnitude depending on the number of quarks within
the mesons or baryons. Over the entire \pt region, both the data and the
model exhibit the same qualitative centrality dependence as observed
for 200 GeV \auau collisions~\cite{star_v2cen}: the more peripheral
the collision, the larger the \vtwo values. Compared to the results
for 200 GeV \auau collisions~\cite{star_v2cen}, the splitting of \ks
and \lam $v_2$($p_T$) is smaller in both the mass ordering region
and the hadron-type dependence region. This indicates smaller
collective flow in \cucu than \auau collisions, which will be seen
more clearly in Section~\ref{sect_system}.

The transverse kinetic energy scaling first observed in \auau
collisions is also tested in Fig.~\ref{v2PtMt}. The results in
Fig.~\ref{v2PtMt}~(a1)-(a3) are re-plotted as a function of the
transverse kinetic energy \ket~in Fig.~\ref{v2PtMt}~(b1)-(b3).
The quantity $m$
denotes the rest mass of a given hadron.
In the low \kets~region, $v_2$($m_T - m$) is a
linearly increasing function and independent of hadron mass. Transverse kinetic
energy scaling holds in the region $m_{T}-m < 0.8$ GeV$/c^2$, as
observed in \auau collisions~\cite{star_v2cen,phenix_scalv2}.
Calculations using ideal hydrodynamics are shown in each panel
as a function of \kets. Contrary to the mass
ordering as a function of $p_T$, the model shows the reversed mass
ordering as a function of \kets: the heavier the hadron, the larger
the $v_2$($m_T - m$) value. The results of ${K}^{0}_{S}$, $\Lambda$ and $\Xi$ exhibit
$m_T - m$ scaling in each centrality bin, while the model does not show any
scaling. Since no pion results are available, the
scaling test of the data is not conclusive.
All these effects can be seen more clearly in Fig.~\ref{v2PtMtZoomin}.

%\clearpage

%\clearpage
\subsection{Quark-number scaling}

In \auau collisions in the intermediate \pt region, 2 $<$ \pt $<$ 4
\GeVc, the baryon-meson grouping of $v_2$($p_T$) follows the
number-of-quark scaling: the $v_{2}$ of all hadrons fall onto a universal curve
once \vtwo and \pt are divided by the number of quarks, $n_q$, in
a given hadron~\cite{62gevv2,star_v2cen}. The observed scaling can
be explained by the coalescence or recombination
models~\cite{Coal_Molnar,reco_Fries,reco_Rudy}, indicating the
constituent quark degree of freedom has been manifested before
hadronization takes place. $n_q$ scaling is tested for various
centrality bins in 200 GeV \cucu collisions: $v_2$($p_T$) and
$v_2$($m_T - m$) scaled by $n_q$ are shown in Fig.~\ref{v2NqPtMt}~(a1)-(a3)
and (b1)-(b3), respectively.
The $n_{q}$-scaling formula of Ref.~\cite{xinv2}, which can be written as
\begin{equation}
\frac{f_{v_{2}}(p_{T})}{n_{q}} = \frac{a}{1+e^{-(p_{T}/n_{q}-b)/c}}-d
\label{Equation:Fitv2}
\end{equation}
has been fitted to the data both in \pt and \ket for each
centrality bin.
In this formula, $a$, $b$, $c$ and $d$ are fit parameters, $n_{q}$ is the
number of quarks. The $n_q$ scaling is observed
for $p_T$/$n_q$ $>$ 0.8 \GeVc, whereas it is seen for the entire
($m_{T} - m$)/$n_q$ region.
Below ($m_{T} - m$)/$n_q$ \sims 0.4
GeV$/c^2$, the \ket~scaling which was established at low $p_{T}$, now scaled by $n_{q}$ leads
to the combined ($m_{T} - m$)/$n_q$ scaling.
The universal $n_q$ scaling of $v_{2}$ suggests the manifestation of
early partonic dynamics in both \auau and \cucu
collisions.

\subsection{Centrality and system-size dependence}
\label{sect_system}

\begin{table*}[ht]
\centering
\caption{Participant eccentricity \eparttwos~and number of participants \npart~from the
Monte Carlo Glauber
model~\cite{ms03,glauber} and Color Glass Condensate (CGC)
model~\cite{CGCa,CGCb,CGCc, CGCd} calculations in \auau and \cucu collisions at
\sqrtsNN = 200 GeV. The quoted errors are total statistical and systematic
uncertainties added in quadrature.}
\begin{tabular}{ccccc} \hline\hline
&Centrality&\eparttwos (CGC)&\eparttwos (Glauber)&\npart\\ \hline
     \auau&\ $0-80$\%\  &$0.338\pm0.002$& $0.302\pm0.004$&$126\pm8$ \\
      ~&\ $0-10$\%\  &$0.148\pm0.001$& $0.123\pm0.003$&$326\pm6$ \\
      ~&\ $10-40$\%\  &$0.353\pm0.001$&$0.296\pm0.009$&$173\pm10$ \\
      ~&\ $40-80$\%\  &$0.554\pm0.002$&$0.533\pm0.018$&$42\pm7$ \\
      \cucu&\ $0-60$\%\  &$0.336\pm0.009$& $0.350\pm0.008$&$51\pm2$ \\
      ~&\ $0-20$\%\  &$0.230\pm0.010$& $0.235\pm0.008$&$87\pm2$ \\
      ~&\ $20-60$\%\  &$0.434\pm0.003$& $0.468\pm0.016$&$34\pm1$ \\
      ~&\ $0-10$\%\  &$0.187\pm0.002$& $0.197\pm0.002$&$99\pm2$ \\
      ~&\ $10-20$\%\  &$0.281\pm0.002$& $0.279\pm0.008$&$75\pm2$ \\
      ~&\ $20-30$\%\  &$0.360\pm0.003$& $0.369\pm0.009$&$54\pm1$ \\
      ~&\ $30-40$\%\  &$0.428\pm0.002$& $0.458\pm0.017$&$38\pm1$ \\
      ~&\ $40-50$\%\  &$0.490\pm0.002$& $0.550\pm0.021$&$26\pm1$ \\
      ~&\ $50-60$\%\  &$0.555\pm0.004$& $0.643\pm0.031$&$17\pm1$ \\ \hline
      \hline
\end{tabular}

\label{tab:ecc}
\end{table*}

\begin{figure*}[ht]
\includegraphics[width=0.7\textwidth]{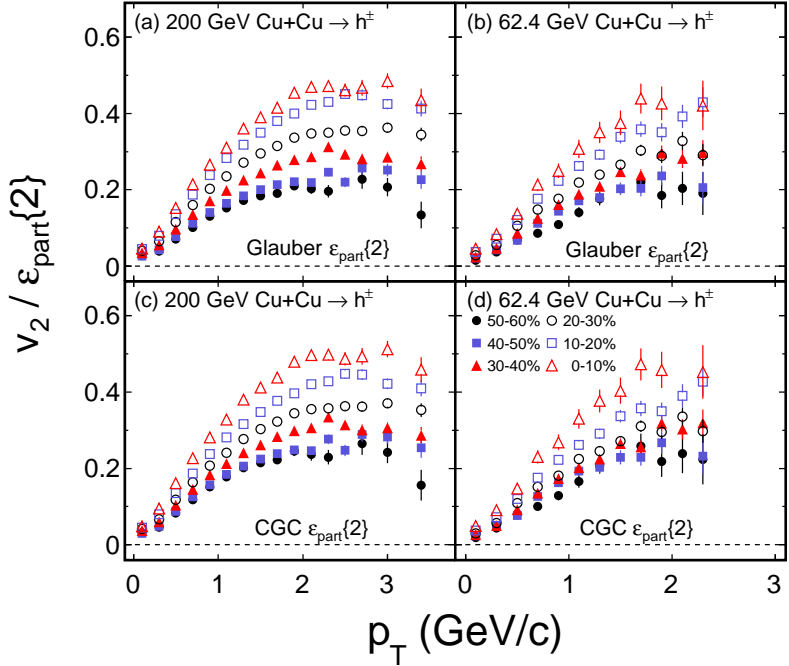}
\caption{ (Color online) $v_2$ scaled by
 participant eccentricity as a function of $p_T$ in \sqrtsNN =
 200 and 62.4 GeV \cucu collisions. The error bars are shown only for the statistical uncertainties.} \label{v2Eccchgpt}
\end{figure*}

\begin{figure*}[ht] \vskip 0cm
\includegraphics[width=0.7\textwidth]{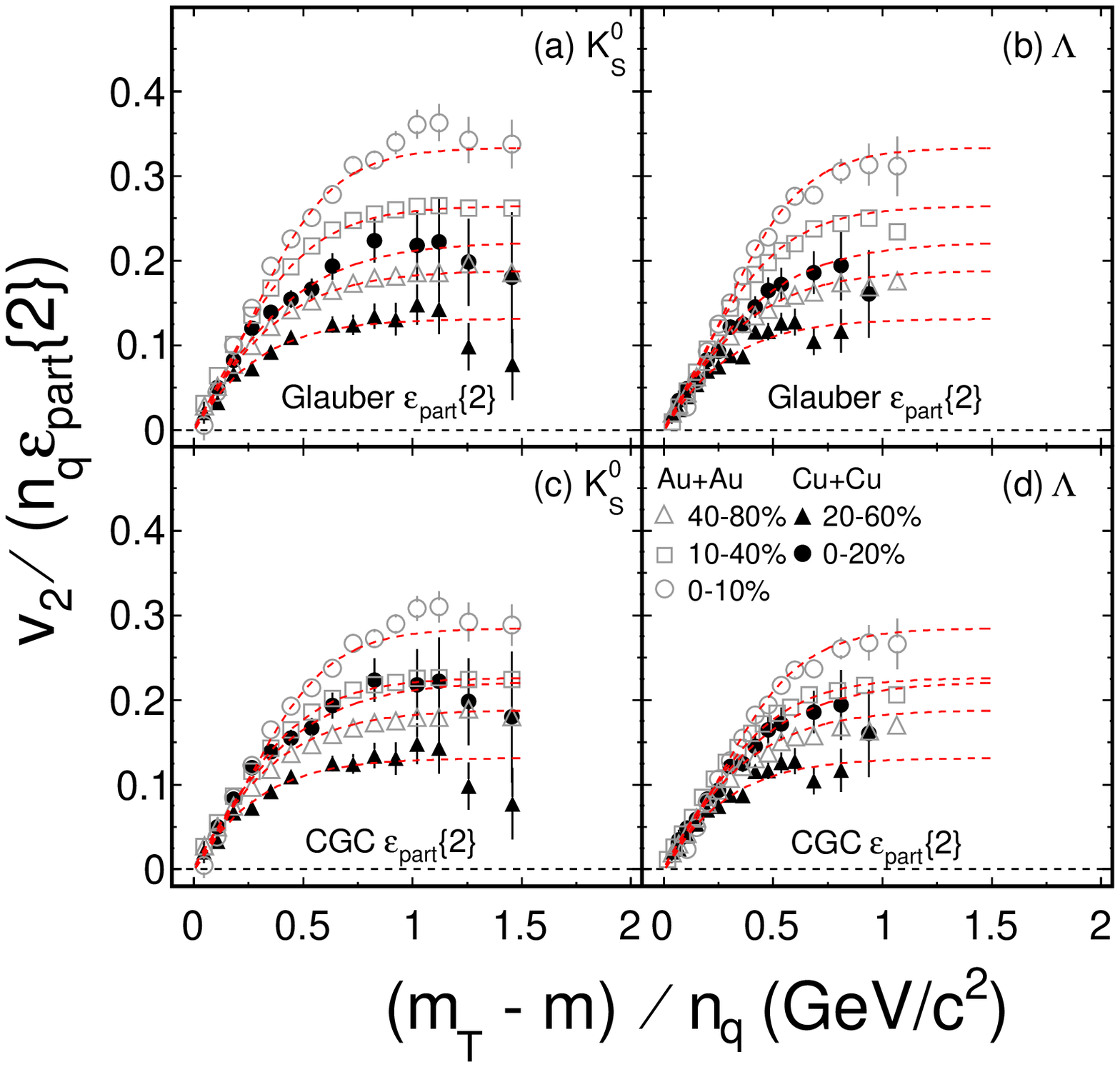}
 \caption{ (Color online) Centrality dependence of \vtwo scaled by
 number of quarks and participant eccentricity
($v_2/$($n_q$\eparttwos)) for \ks(left) and
\lam(right) as a function of ($m_T - m$)/$n_q$ in $0-10$\%, $10-40$\% and $40-80$\% \auau collisions (open symbols)~\cite{star_v2cen}
and $0-20$\% and $20-60$\% \cucu collisions (solid symbols) at \sqrtsNN
= 200 GeV. Curves
are the results of $n_q$-scaling fits from Eq.~(\ref{Equation:Fitv2}) normalized by \eparttwo to combined \ks and \lam for five centrality bins.
 At a given $p_{T}$, from top to bottom, the curves show a decreasing trend as \npart~decreases.
The error bars are shown only for the statistical uncertainties.} \label{nqecc_centrality}
\end{figure*}

\begin{figure*}[ht]
\vskip 0cm
\includegraphics[width=0.8\textwidth]{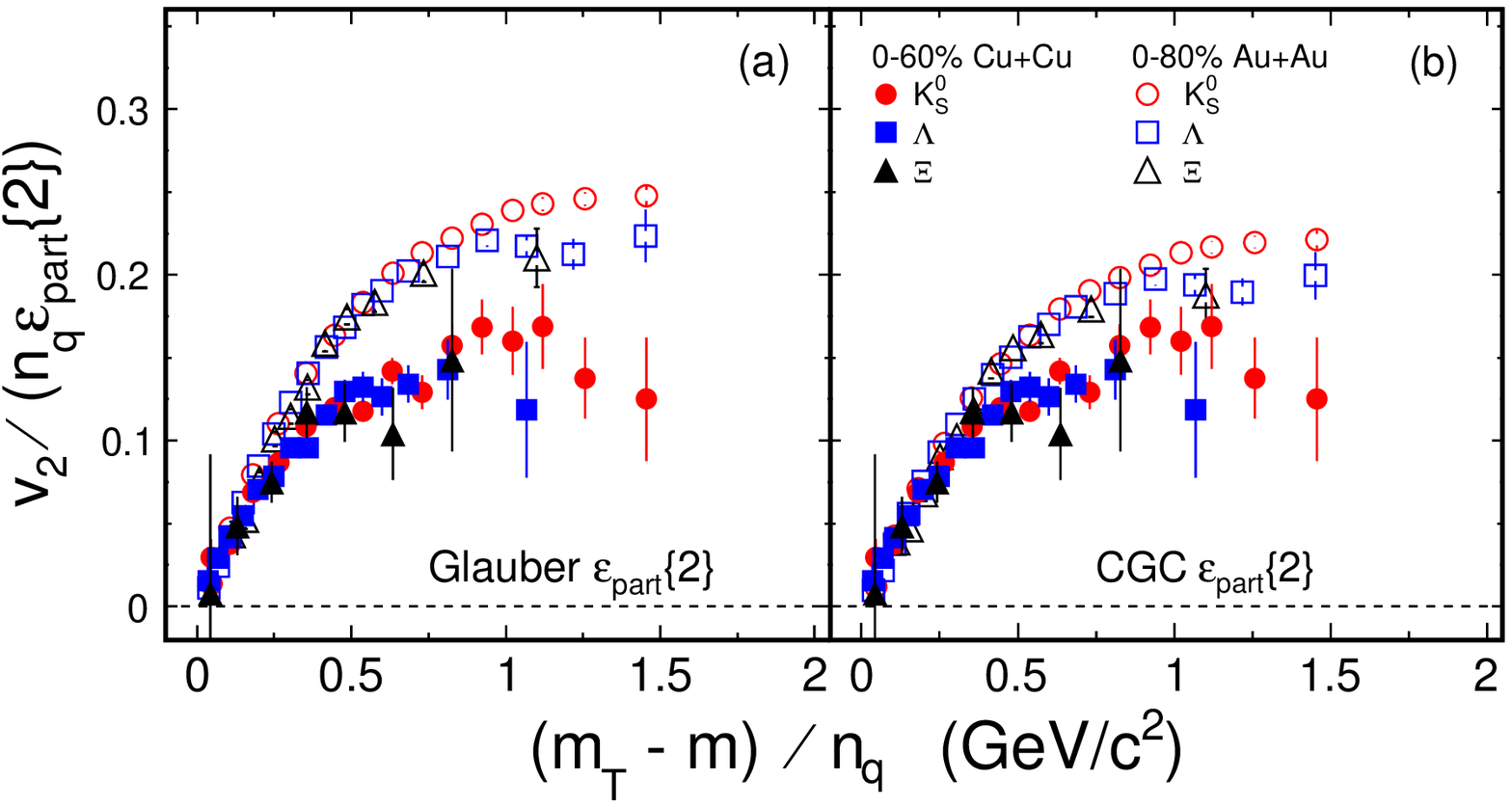}
\caption{ (Color online) Number of quarks and participant
eccentricity scaled \vtwo ($v_2/$($n_q$\eparttwos)) of
identified particles as a function of ($m_T - m$)/$n_q$ in $0-80$\% \auau collisions (open symbols)~\cite{star_v2cen}
and $0-60$\% \cucu collisions (closed symbols) at \sqrtsNN
= 200 GeV. Circles, squares and
triangles represent the data for \kss, $\Lambda$ and $\Xi$,
respectively. The error bars are shown only for the statistical uncertainties.} \label{nqecc_minibias}
\end{figure*}

The centrality and system-size dependence of \vtwos~is related to the physics
of the system created in high energy nuclear collisions.
In the ideal hydrodynamic limit the centrality dependence of elliptic flow is mostly defined by
the elliptic anisotropy of the overlapping region of the colliding nuclei,
and in the low-density limit by the product of the
elliptic anisotropy and the multiplicity.
Thus, the centrality and system-size dependence of elliptic flow should be a good indicator of
the degree of equilibration reached in the reaction~\cite{centdependence}.

For a study of the centrality dependence of $v_2$($p_T$)
in \cucu collisions together with \auau collisions, we divide
$v_2$($p_T$) by the initial spatial anisotropy,
eccentricity, to remove this geometric effect.
The participant eccentricity is the initial configuration
 space eccentricity of the participants which is defined by~\cite{partecc}
 \begin{equation}
\varepsilon_{\rm part} = \frac{\sqrt{(\sigma_{y}^{2} - \sigma_{x}^{2})+4(\sigma_{xy}^{2})}}{\sigma_{y}^{2} + \sigma_{x}^{2}}
\label{Equation:partiecc}
\end{equation}
In this formula, $\sigma_{x}^{2} = \langle x^{2} \rangle - \langle x\rangle^{2}$,
$\sigma_{y}^{2} = \langle y^{2} \rangle - \langle y\rangle^{2}$ and
 $\sigma_{xy} = \langle xy \rangle - \langle x \rangle\langle y \rangle$,
 with $x$, $y$ being the position of the participating nucleons in the transverse
 plane.
 The root mean square of the participant
eccentricity
\begin{equation}
\varepsilon_{\rm part}\{2\} = \sqrt{\langle\varepsilon_{\rm part}^{2}\rangle}
\label{Equation:epart2}
\end{equation}
is calculated from the Monte Carlo Glauber
model~\cite{ms03,glauber} and Color Glass Condensate (CGC)
model~\cite{CGCa,CGCb,CGCc, CGCd}. (See Table~\ref{tab:ecc} for \eparttwos.)
Since the FTPC event plane is constructed from the hadrons which have their
origin in participant nucleons and the FTPC event plane resolution is less
than 0.2, what we actually measure is
the root mean square of $v_{2}$ with respect to the participant plane
~\cite{epart21}. In this case,
\eparttwos~is the appropriate measure
of the initial geometric anisotropy taking the event-by-event
fluctuations into account~\cite{eparttwo, epart2, epart21}.
Figure~\ref{v2Eccchgpt} shows the centrality dependence of
$v_2$($p_T$)/\eparttwos~for \chg in
200 and 62.4 GeV \cucu collisions. For a given centrality bin,
$v_2$($p_T$)/\eparttwos~initially increases with \pt and then
flattens or falls off at higher $p_T$. After the geometric effect is
removed, the ordering of the distributions as a function of
centrality, observed in Fig.~\ref{v2chgpt}, is reversed: the more central the collision, the higher
the $v_2$($p_T$)/\eparttwos. This suggests that the
strength of collective motion is larger in more central collisions.

To further study the centrality
dependence of strange hadron $v_2$, we normalized the $n_q$-scaled
values by \eparttwos~and plotted them as a function of (\kets)/$n_q$.
 The centrality dependence of \ks and $\Lambda$
results are shown in Fig.~\ref{nqecc_centrality}.
The full symbols show from top to bottom the results
from $0-20$\% and $20-60$\% centrality \cucu collisions. For comparison,
the results from 200 GeV \auau collisions~\cite{star_v2cen} are
shown by open symbols in Fig.~\ref{nqecc_centrality}.
The results in \auau collisions are slightly
different ($\sim10\%$ larger) from the previous published results~\cite{star_v2cen},
which were calculated directly from the wide centrality bins.
From top to bottom, the results are from $0-10$\%,
$10-40$\% and $40-80$\% centrality bins.
For clarity, \ks and $\Lambda$ results are shown in different
panels. Curves represent
$n_{q}$-scaling fits from Eq.~(\ref{Equation:Fitv2}) normalized by \eparttwo to the combined data of \ks and $\Lambda$
for five centrality bins.
For a given centrality, \ks and $\Lambda$ results follow a universal curve,
which means partonic collective flow is explicitly seen in the measured scaling
with $n_q$ and \eparttwos. For a given collision system, the stronger
partonic collective flow is apparent as higher scaled $v_{2}$
value in more central collisions. To study the system-size
dependence of the scaling properties, the results from
 $0-60$\% centrality \cucu and $0-80$\% \auau  collisions are shown in
Fig.~\ref{nqecc_minibias}. The stronger collective motion in \auau
compared to \cucu collisions becomes obvious although the constituent
quark degrees of freedom have been taken into account in both systems.

In the ideal hydrodynamic limit where dynamic thermalization is reached, the
mean free path is much less than the geometric size of the system.
The geometric size of the system and the centrality dependence of flow is totally governed by the
initial geometry (eccentricity)~\cite{centdependence}. As there is
no universal scaling with the eccentricity among either different
collision centralities or different collision system sizes,
this indicates that the ideal hydrodynamic limit is not reached in \cucu
collisions, presumably because the assumption of thermalization is not attained.
In addition, $v_2$/($n_q$\eparttwos) shows an
increasing trend as a function of \npart~(See Fig.~\ref{nqecc_centrality}).
Table~\ref{tab:ecc} lists the values of eccentricity and \npart~for the used
centrality bins in \auau and \cucu collisions.
This suggests that the
measured \vtwo is not only dependent on the initial geometry, but also
on \npart.

Theoretical analyses found that the centrality and system-size dependence of
\vtwo can be described by a simple model based on eccentricity scaling and incomplete
thermalization. Within these models the lack of perfect equilibration allows for estimates of
the effective parton cross section in the quark-gluon plasma and of the shear viscosity
to entropy density ratio ($\eta/s$)~\cite{Drescher07, knudsen}.
 Thus, the \vtwo results from \cucu collisions reported in this paper
should allow extraction of $\eta/s$ and extrapolation to the ideal hydrodynamic limit.

\section{Summary}
\label{sect_summary}

We present STAR results on midrapidity elliptic flow \vtwo for
charged hadrons \chg and strangeness containing hadrons \kss, $\Lambda$, \xii and $\phi$
from \cucu collisions at \sqrtsNN = 62.4 and 200 GeV at RHIC. The
centrality dependence of \vtwo for different system-sizes as a
function of the transverse momentum \pt is presented. To estimate
the systematic uncertainties, we studied various measurement
methods. Below \pt \sims 4 \GeVc, non-flow correlations are reduced
with the event plane constructed from hadrons produced in the
forward regions ($2.5 < |\eta| < 4.0$).
We obtained an estimate
of the systematic uncertainties due to remaining non-flow
contributions based on correlations measured in \pp
collisions.

For a given centrality bin, \pt and hadron-type dependences of strange hadron \vtwo are similar to
those found in \auau collisions~\cite{star_v2cen}: (i) in the
low \pt region, $p_T < 2\ \GeVc$, the hadron mass
hierarchy is observed as expected in ideal hydrodynamic calculations: at
fixed \pts, the larger the hadron mass, the smaller the \vtwos. (ii)
in the intermediate \pt region, $2 < p_T < 4\ \GeVc$, \vtwo as a
function of either \pt or \ket follows a scaling with the number of
constituent quarks $n_q$. Larger $v_2$/($n_q$\eparttwos) values are seen in more
central collisions, indicating stronger collective flow
developed in more central collisions.
The comparison with \auau collisions which go further in density
shows eccentricity scaled $v_2$ values depend on the system size (\npart).
This suggests
that the ideal hydrodynamic limit is not reached in \cucu
collisions, presumably because the assumption of thermalization is not attained.

\section{Acknowledgments}
We thank the RHIC Operations Group and RCF at BNL, the NERSC Center at LBNL and the Open Science Grid consortium for providing resources and support. This work was supported in part by the Offices of NP and HEP within the U.S. DOE Office of Science, the U.S. NSF, the Sloan Foundation, the DFG cluster of excellence `Origin and Structure of the Universe' of Germany, CNRS/IN2P3, STFC and EPSRC of the United Kingdom, FAPESP CNPq of Brazil, Ministry of Ed. and Sci. of the Russian Federation, NNSFC, CAS, MoST, and MoE of China, GA and MSMT of the Czech Republic, FOM and NWO of the Netherlands, DAE, DST, and CSIR of India, Polish Ministry of Sci. and Higher Ed., Korea Research Foundation, Ministry of Sci., Ed. and Sports of the Rep. Of Croatia, Russian Ministry of Sci. and Tech, and RosAtom of Russia.

%
%\clearpage
%--=================================================================

%
%%--==========================================================================
\end{document}